\documentclass{rmf-d}
\usepackage{nopageno,rmfbib,multicol,times,epsf,amsmath,amssymb,cite,float}
\usepackage[latin1]{inputenc}
\usepackage{caption2}
\usepackage{graphicx}
\usepackage{gensymb}
\usepackage{hyperref}

\def\rmfcornisa{Research OR Education of Physics \hfill\rmf\ {\bf ??} (*?*) ???--???
\hfill MES? A\~NO?}

\def\rmfcintilla{{\it Rev.\ Mex.\ Fis.\/} {\bf ??} (*?*) (????) ???--???}
\clearpage \rmfcaptionstyle 
\begin{document}
\title{   Light mesons with one dynamical gluon within basis light-front quantization
\vspace{-6pt}}
\author{ Jiangshan Lan$^{a,b,c,d~*}$, Kaiyu Fu$^{a,b,c}$, Chandan Mondal$^{a,b,c}$, Xingbo Zhao$^{a,b,c}$, James P. Vary$^e$    }
\address{{\rm (BLFQ Collaboration)}\\
$^a$ Institute of Modern Physics, Chinese Academy of Sciences, Lanzhou 730000, China \\
$^b$ CAS Key Laboratory of High Precision Nuclear Spectroscopy, Institute of Modern Physics, Chinese Academy of Sciences, Lanzhou 730000, China\\
$^c$ School of Nuclear Science and Technology, University of Chinese Academy of Sciences, Beijing 100049, China\\
$^d$ Lanzhou University, Lanzhou 730000, China\\
$^e$ Department of Physics and Astronomy, Iowa State University, Ames, Iowa 50011, USA}
%\collaboration{BLFQ Collaboration}

\maketitle
\recibido{day month year}{day month year
\vspace{-12pt}}
\begin{abstract}
\vspace{1em} 
%We study the light meson with one dynamical gluon on the light-front quantum chromodynamics (QCD) Hamiltonian as well as a three-dimensional confinement. After fitting the light meson mass spectroscopy,  the light-front wave function provides a good description of the pion electromagnetic form factor, and the valence quark distribution functions following QCD scale evolution.  
We investigate the structure of the light meson with one dynamical gluon from the light-front QCD Hamiltonian, determined for their constituent quark-antiquark and quark-antiquark-gluon Fock sectors, together with a three-dimensional confinement. After fitting the light meson mass spectroscopy, the light-front wave functions provide a good quality description of the pion electromagnetic form factor, and the valence quark distribution functions following QCD scale evolution.
 \vspace{1em}
\end{abstract}
\keys{  Basis light-front quantization; Light mesons; Mass spectroscopy; Light-front wavefunctions; Parton distribution functions \vspace{-4pt}}
\pacs{   \bf{\textit{14.40.Be, 12.38.Aw, 33.15.Ta, 13.40.Gp}}    \vspace{-4pt}}
\begin{multicols}{2}

%%%%%%%%%%%%%%%%%%%%%%%%%%%%%%%%%%%%%%%%%%%%%%%%%%%%%%%%%%%%%%%%%%

\section{Introduction}

Basis Light-Front Quantization (BLFQ) is a Hamiltonian formalism that incorporates the advantages of the light-front dynamics and {\it ab initio} nuclear structure methods, which has been developed for solving many-body bound state problems in quantum field theories~\cite{Vary:2009gt,Wiecki:2014ola,Li:2015zda,Zhao:2014xaa}.
%It has been successfully applied to the single electron problem and the strong coupling bound-state positronium problem in quantum electrodynamics (QED) systems. Meanwhile, i
This approach has been successfully applied to QCD  systems, such as, heavy mesons~\cite{Li:2017mlw,Tang:2018myz,Chen:2018vdw,Li:2019kpr,Lan:2019img}, light mesons~\cite{Jia:2018ary,Lan:2019vui,Lan:2019rba}, and baryon~\cite{Mondal:2019jdg,Xu:2021wwj}. Recently, it has been employed for the first time to investigate light mesons with one dynamical gluon from the QCD Hamiltonian, which includes a vector coupling vertex of quark-gluon interaction and an instantaneous gluon exchange interaction~\cite{Lan:2021wok}. 
% In this work, we consider an effective light-front Hamiltonian and solve for its mass eigenvalues and eigenstates at the scales suitable for low-resolution probes. 
In this work, we present the electromagnetic form factor (EMFF) and the  parton distribution functions (PDFs) of the pion with one dynamical gluon from the resulting light-front wave functions (LFWFs) obtained as eigenvectors of this Hamiltonian. 

The EMFF can be expressed in terms of the meson LFWFs using the Drell-Yan-West formula \cite{Brodsky:2007hb},
\begin{align}
F(Q^2)=&\sum_f e_f \sum_{\mathcal{N},\,\lambda_i} \int_{\mathcal{N}}  \Psi^{\mathcal{N},\,M_J=0\,*}_{\{x_i,\vec{p}^{\,\prime}_{\perp i},\lambda_i\}}\,\Psi^{\mathcal{N},\,M_J=0}_{\{x_i,\vec{p}_{\perp i},\lambda_i\}}\,,
\label{eqn:ff_q}
\end{align}
where $\int_{\mathcal{N}} \equiv  \prod_{i=1}^\mathcal{N} \int \left[\frac{dx\,d^2\vec{p}_\perp}{16\pi^3}\right]_i 16\pi^3 \delta(1-\sum x_j)\delta^2(\sum \vec{p}_{\perp}^ j)$ and
for a struck parton, $\vec{p}^{\,\prime}_{\perp i}=\vec{p}_{\perp i}+ (1-x_i)\vec{q}_\perp$, while $\vec{p}^{\,\prime}_{\perp i}=\vec{p}_{\perp i}- x_i \vec{q}_\perp$ for the spectators. Considering the frame, where $q^+=0$, $Q^2=-q^2=\vec{q}^{\,2}_{\perp}$. The electric charge $e_u(e_{\bar{d}})=\frac{2}{3}(\frac{1}{3})$, while gluon is chargeless.  Here $\psi^{\mathcal{N}=2}(\{\overline{\alpha}_i\})$ and $\psi^{\mathcal{N}=3}(\{\overline{\alpha}_i\})$ are the LFWFs of the Fock sectors $|q\bar{q}\rangle$ and $|q\bar{q}g\rangle$, respectively, in the BLFQ basis obtained from diagonalizing the mass-squared operator. 

The PDF is the probability of finding a collinear parton (quark and gluon) carrying the longitudinal momentum fraction $x$ inside the hadron. Based on our LFWFs, the valence quark (antiquark) and the gluon PDFs in the pion are given by
\begin{align}
f_i(x)=\sum_{\mathcal{N},\,\lambda_i} \int_{\mathcal{N}}  \Psi^{\mathcal{N},\,M_J=0\,*}_{\{x_i,\vec{p}_{\perp i},\lambda_i\}}\,\Psi^{\mathcal{N},\,M_J=0}_{\{x_i,\vec{p}_{\perp i},\lambda_i\}}\,\delta(x-x_i)\,,
\label{eqn:pdf}
\end{align}
where $i=q,\bar{q},g$ labels the valence quark, valence antiquark, and gluon, respectively. At the model scale the PDFs for the valence quark (antiquark) are normalized as 
\begin{align}
\int_0^1 f_{q/\bar{q}} (x)dx=1, 
\end{align}
and those PDFs together with the gluon PDF satisfy the momentum sum rule: 
\begin{align}
\int_0^1 \sum_i x f_i(x) dx=1.
\end{align}

%%%%%%%%%%%%%%%%%%%%%%%%%%%%%%%%%%%%%%%%%%%%%%%%%%%%%%%%%%%%%%%%%%

\section{Light-front Hamiltonian}

In the BLFQ approach, one obtains the LFWF by solving an eigenvalue problem of the Hamiltonian: 
\begin{align}
P^-P^+|{\Psi}\rangle=M^2|{\Psi}\rangle,
\end{align}
 where $P^\pm=P^0 \pm P^3$ employs the light-front (LF) Hamitonian ($P^-$) and the longitudinal momentum ($P^+$) of the system, respectively, with the mass squared eigenvalue $M^2$.

\begin{figure}[H]
    \centering
    \includegraphics[width=0.9\linewidth]{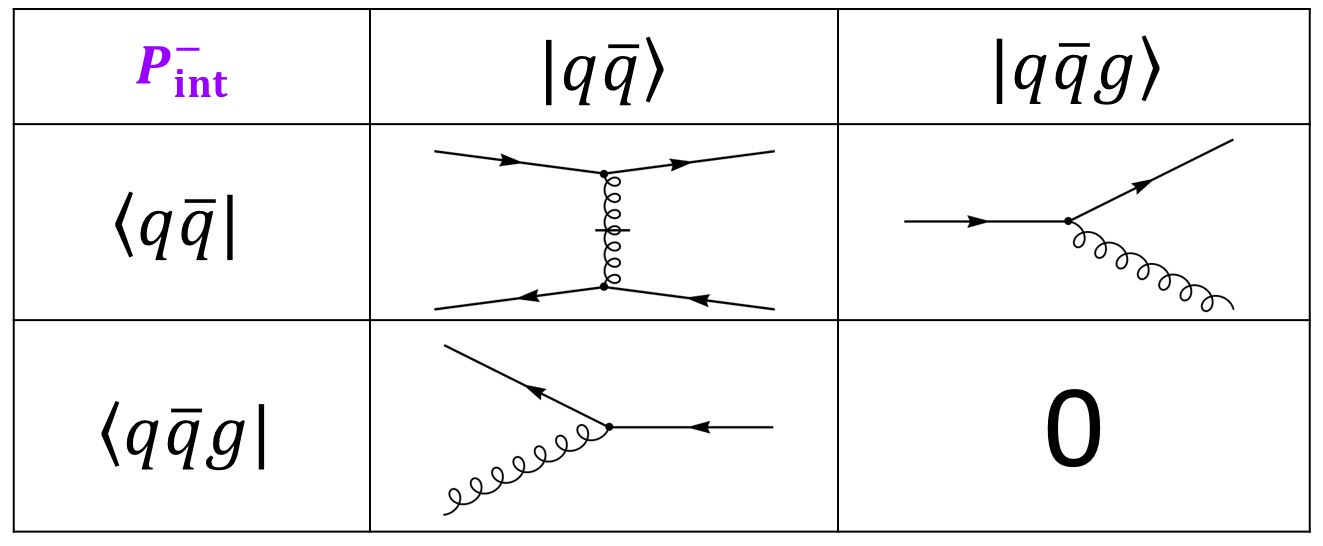}
    \caption{The interaction between quark and gluon on light front.
}
    \label{hint}
\end{figure}

\begin{figure}[H]
    \centering
    \includegraphics[width=\linewidth]{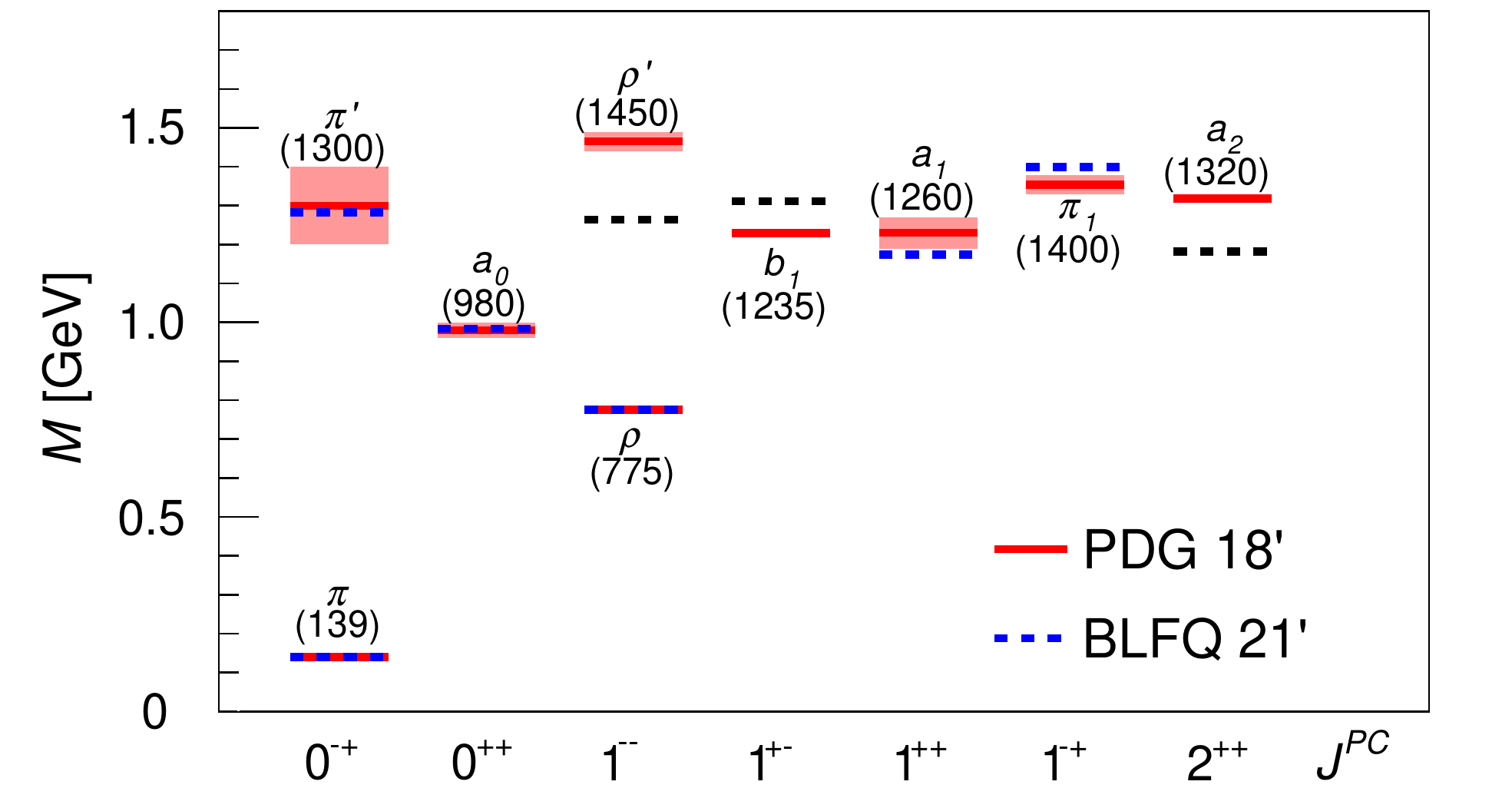}
    \caption{The mass spectra of unflavored light mesons. Our results (dashed bars) are compared with the experimental data taken from PDG (red-solid bars)  \cite{Tanabashi:2018oca}.}
    \label{mass}
\end{figure}

 At fixed LF time ($x^+=t+z$), the meson state can be expressed in terms of various quark ($q$), antiquark ($\bar q$) and gluon $(g)$ Fock components,
\begin{align}\label{Eq1}
|\Psi\rangle=\psi_{(q\bar{q})}|q\bar{q}\rangle+\psi_{(q\bar{q}g)}|q\bar{q}g\rangle+\dots\, , 
\end{align}
where the LFWFs $\psi_{(\dots)}$ correspond to the probability amplitudes  to find different parton configurations in the meson. At the initial scale where the mesons are described by $|q\bar{q}\rangle$ and $|q\bar{q}g\rangle$, 
we adopt the LF Hamiltonian~\cite{Lan:2021wok} 
\begin{align}
P^-= P^-_{\rm KE} +P^-_{\rm C} + P^-_{\rm int},\label{hami}
\end{align}
 where $P^-_{\rm KE}$ is the kinetic energy of the quark and gluon, $P^-_{C}$ represents the confining potential on both the transverse and the longitudinal directions~\cite{Brodsky:2014yha,Li:2015zda}, and the $P^-_{\rm int}$ corresponds to the interaction between quark and gluon on LF~\cite{Brodsky:1997de} as illustrated in Figure~\ref{hint} .
With the framework of BLFQ, we consider the effective LF Hamiltonian, Eq.~(\ref{hami}), and solve for its mass eigenvalues and eigenstates at the scales suitable
for low-resolution probes. We fix the model parameters by fitting the known masses of $\pi$, $\rho$, $a_0$, $a_1$, $\pi'$, and $\pi_1$ as shown in Fig.~\ref{mass}.
% , the model parameters are fixed, then the LFWFs are obtained~\cite{Lan:2021wok} .  

%%%%%%%%%%%%%%%%%%%%%%%%%%%%%%%%%%%%%%%%%%%%%%%%%%%%%%%%%%%%%%%%%%
\section{Numerical results}
\begin{figure*}
    \centering
    \includegraphics[width=0.8\linewidth]{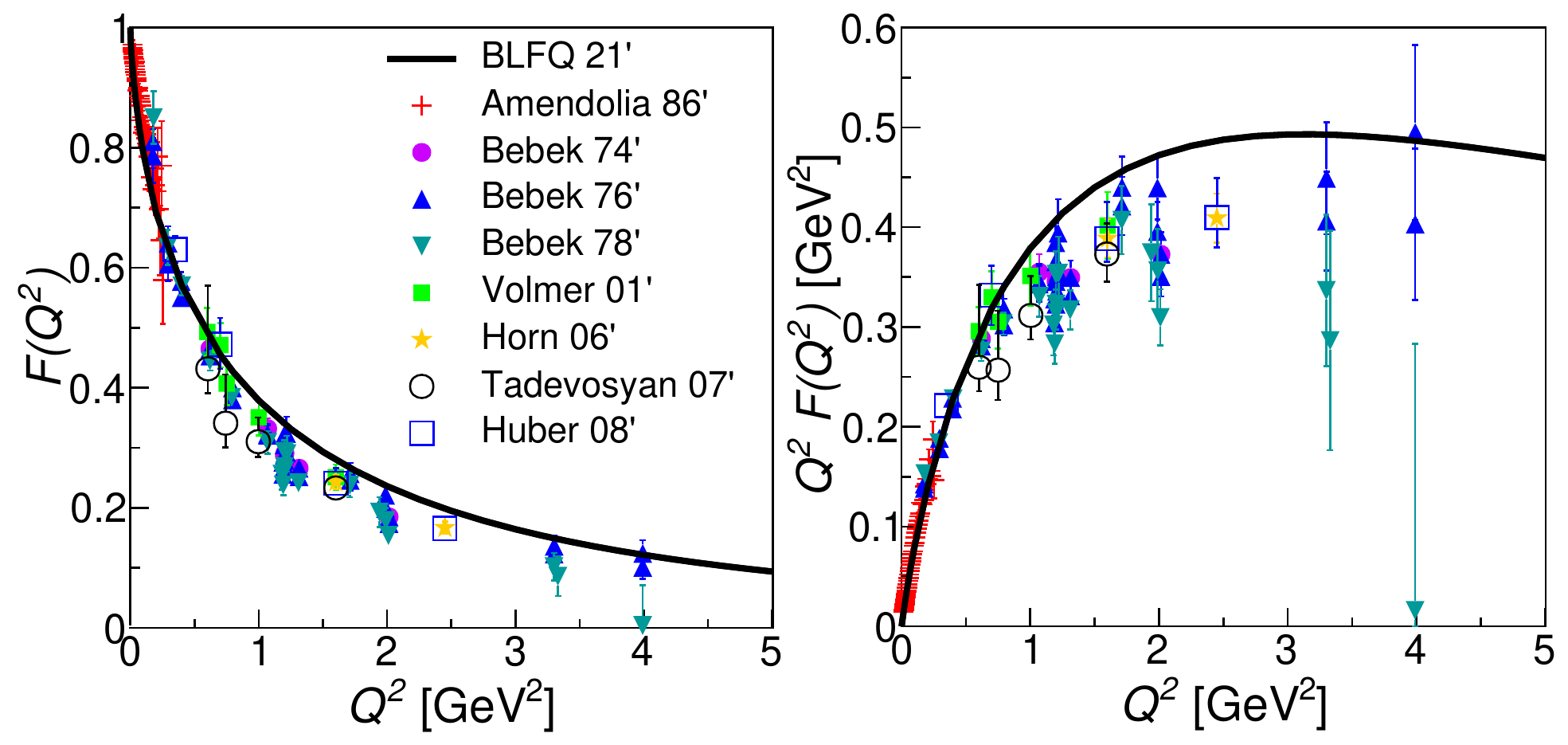}
    \caption{The EMFF of the pion. The data are taken from Refs. \cite{NA7:1986vav,Bebek:1974iz,Bebek:1974ww,Bebek:1977pe,JeffersonLabFpi:2000nlc,JeffersonLabFpi-2:2006ysh,JeffersonLabFpi:2007vir,JeffersonLab:2008jve}.}
    \label{ff}
\end{figure*}

We employ our resulting LFWFs to compute the pion EMFF. In Fig.~\ref{ff}, our prediction for the EMFF of the charged pion is compared to the experimental data~\cite{NA7:1986vav,Bebek:1974iz,Bebek:1974ww,Bebek:1977pe,JeffersonLabFpi:2000nlc,JeffersonLabFpi-2:2006ysh,JeffersonLabFpi:2007vir,JeffersonLab:2008jve}. We find an impressive agreement between our results and the precise low $Q^2$ EMFF data.
%Considering the ultra-violet (UV) regulator $\Lambda_{\rm UV} \sim 1$ GeV, we expectate that our predictions are reliable in the low $Q^2$ regime.

%Look at the left panel of Figure~\ref{ff}, we find that the $F(Q^2) \propto 1/Q^2 $ for large $Q^2$. 

We obtain the pion PDFs at the model scale using our resulting LFWFs in Eq.~(\ref{eqn:pdf}). We then evolve our initial PDFs~\cite{Lan:2021wok} from the model scale ($\mu_0^2$) to a higher scale ($\mu^2$) by the Dokshitzer-Gribov-Lipatov-Altarelli-Parisi (DGLAP) equations \cite{Dokshitzer:1977sg,Gribov:1972ri,Altarelli:1977zs}. We solve the next-to-next-to-leading order~(NNLO) DGLAP equations numerically using the higher order perturbative parton evolution toolkit~\cite{Salam:2008qg}.
The initial scale $\mu_0^2$ is determined by requiring the result after evolution to produce the total first moments of the valence quark PDFs from the global QCD analysis, $\langle x \rangle_{\rm valence}=0.48\pm0.01$ at $\mu^2=5$ GeV$^2$~\cite{Barry:2018ort}. This results in $\mu_0^2=0.340 \pm 0.034$ GeV$^2$.
%, including 10 $\%$ uncertainty. 
We then evolve our initial PDFs to the relevant experimental scale $\mu^2= 16~ {\rm GeV}^2$.  
 
\begin{figure*}
    \centering
    \includegraphics[width=0.485\linewidth]{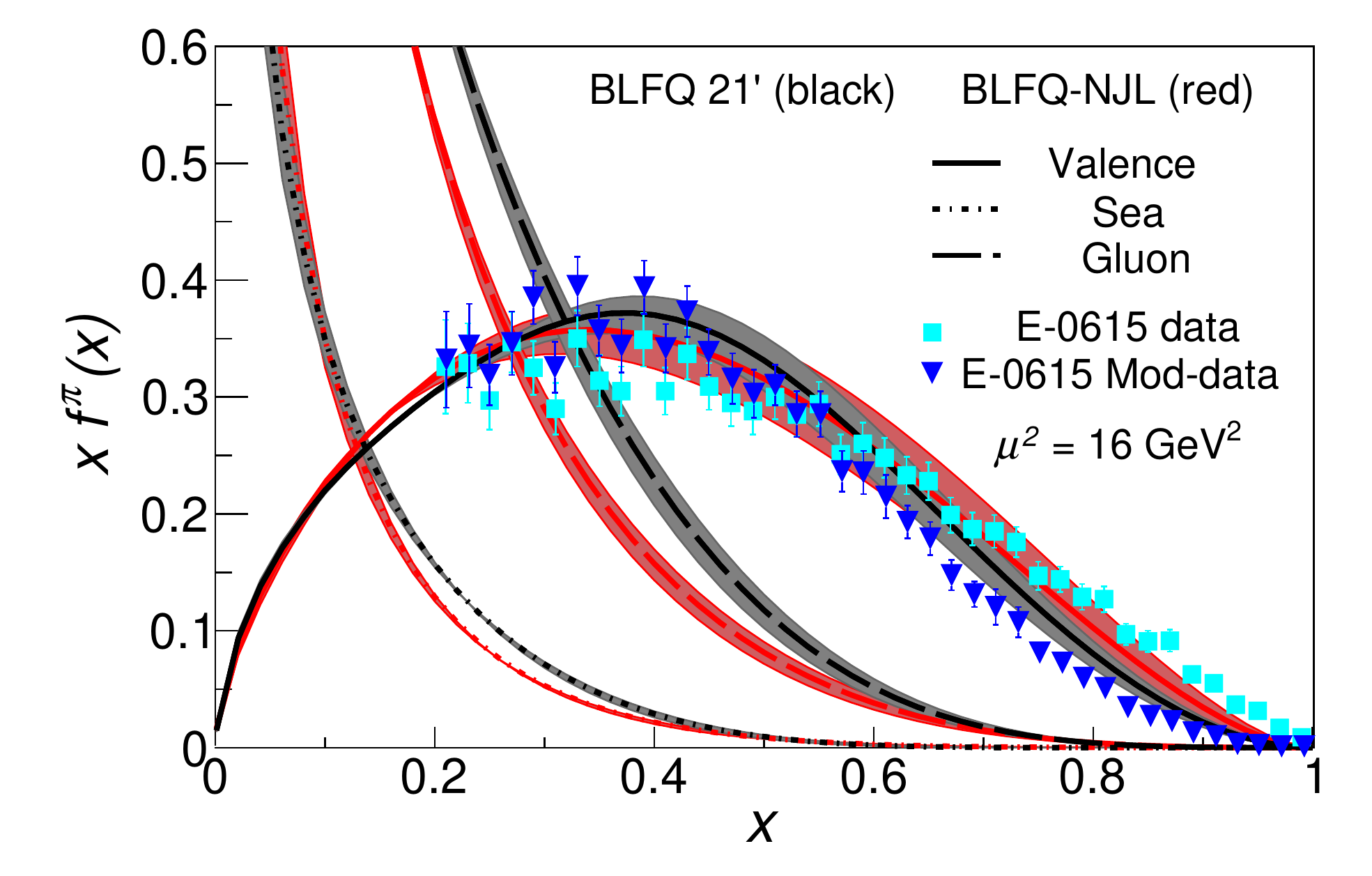}
    \includegraphics[width=0.485\linewidth]{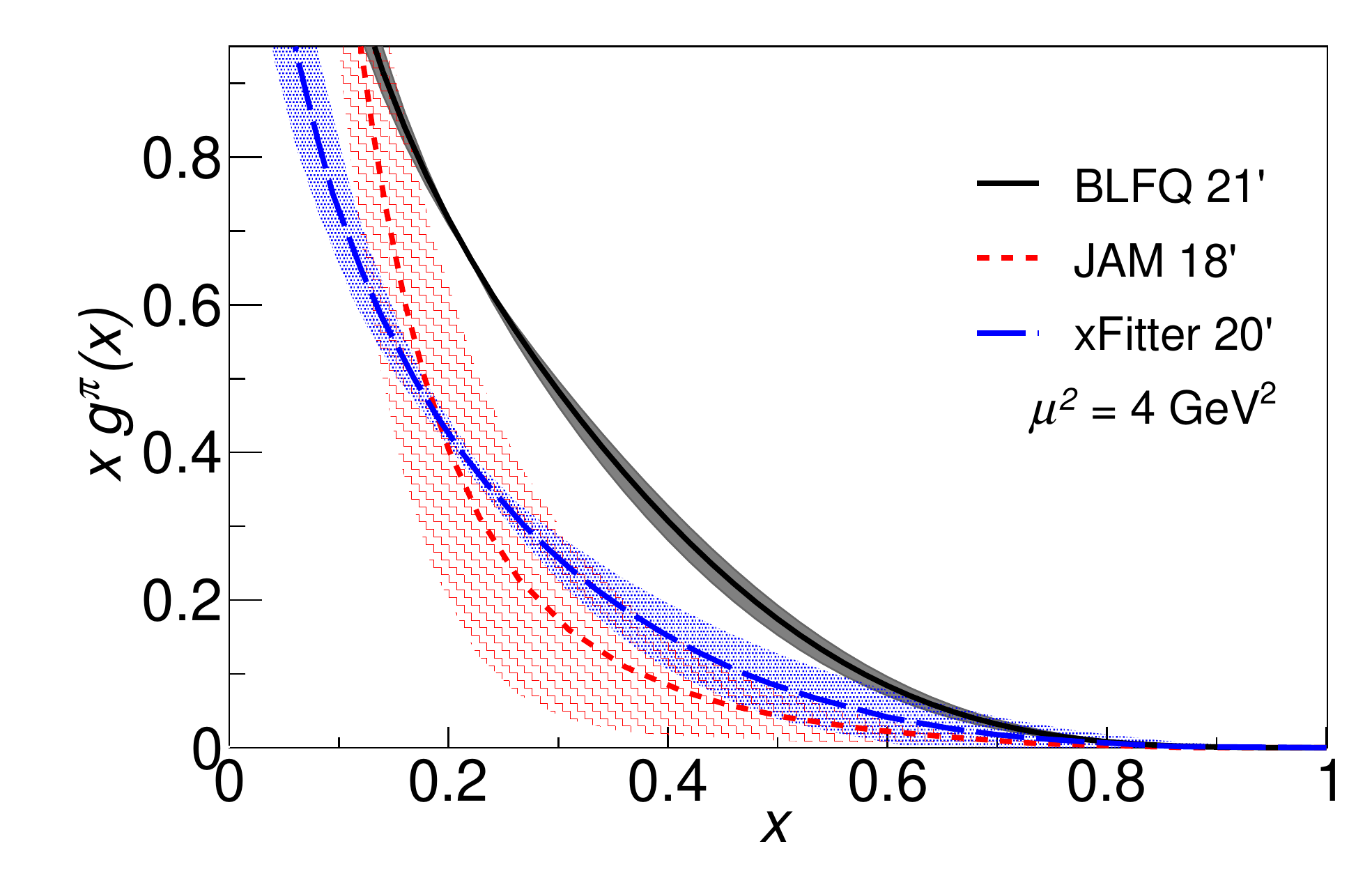}
    \caption{Left panel: the PDFs of the pion at $\mu^2=16$ GeV$^2$. The valence distribution is compared with the original analysis of the FNAL-E615 experiment~\cite{Conway:1989fs} data and with its reanalysis (E615 Mod-data)~\cite{Chen:2016sno}  Right panel: the pion's  gluon PDF at $\mu^2=4$ GeV$^2$ is compared with the global fits, JAM~\cite{Barry:2018ort} and xFitter~\cite{Novikov:2020snp}. }
    \label{pdf}
\end{figure*}

In Fig.~\ref{pdf}, we show the valence quark, sea quark, and gluon PDFs of the pion. 
The black lines are our results evolved from the initial scale $(0.340\pm 0.034~\mathrm{GeV}^2)$ using the NNLO DGLAP equations to the experimental scale of $16~\mathrm{GeV}^2$. 
The red lines correspond to BLFQ-NJL predictions~\cite{Lan:2019vui} with the initial scale $(0.240\pm 0.024~\mathrm{GeV}^2)$. Results are compared with the original analysis of the FNAL-E615 experiment~\cite{Conway:1989fs} data and with its reanalysis (E615 Mod-data)~\cite{Chen:2016sno}.
After including one dynamical gluon, we find that the initial scale naturally increases, and the behavior of the pion valence PDF at large $x$, $(1-x)^{1.77}$, approaches the reanalysis of E615 data. We obtain the first moments of the valence quark, sea quark, and gluon distributions at $4$ GeV$^2$, $\langle x\rangle=\{0.483,0.096,0.421\}$, respectively.

The gluon distribution significantly increases in our approach compared to that in the BLFQ-NJL model~\cite{Lan:2019vui,Lan:2019rba} as well as to the global fits~\cite{Barry:2018ort,Novikov:2020snp} as can be noticed from Fig.~\ref{pdf}. We notice that gluons carry $\{39.5,\,42.1,\,43.9,\,44.6,\,45.1\}\%$ of pion momentum at the scale $\mu^2=\{1.69,\,4,\,10,\,16,\,27\}$ GeV$^2$, respectively. 

%Note that our gluon distribution significantly increases compared to the BLFQ-NJL model result. Where the BLFQ-NJL model is based on the pion valence Fock sector and gluons are generated by the scale evolution. While, this work includes a dynamical gluon at the initial scale and gives more gluon information in a larger at large-$x~ (>0.2)$ after scale evolution. 
%Furthermore, the various valence quarks, sea quarks, and gluon contribution to the first moment of the pion at 4 GeV$^2$ are $\{0.483,0.096,0.421\}$. Our result is also supported by a recently DSE model~\cite{Freese:2021zne}. 
%We also find that gluons carry $\{39.5,\,43.9,\,44.6,\,45.1\}\%$ of pion momentum at $\{1.69,\,10,\,16,\,27\}$ GeV$^2$, respectively.


\section{Conclusion and outlook}
We have  investigated the light mesons from the LF QCD Hamiltonian by considering them within the constituent quark-antiquark and the quark-antiquark-gluon Fock components. Together with a three-dimensional confinement in the leading Fock sector, the eigenvalues of the Hamiltonian in BLFQ provide a good quality description of the light mesons' mass spectra. The LFWFs have been employed to compute the pion EMFF and the PDFs. We have obtained excellent agreement with the experimental results in the low-$Q^2$ regime for the pion EMFF. The valence quark PDFs, after DGLAP evolution, are also good agreement with the experimental data. 

The resulting LFWFs for light mesons can be employed to study the generalized parton distributions (GPDs), the transverse momentum dependent parton distributions (TMDs) as well as the double parton correlations (DPDs) etc., in the future. On the other hand, this work can be extended to higher Fock sectors to incorporate, such as, sea degrees of freedom as well.

\section*{Acknowledgements}
C. M. thanks the Chinese Academy of Sciences President's International Fellowship Initiative for the support via Grants No. 2021PM0023. C. M. is supported by new faculty start up funding by the Institute of Modern Physics, Chinese Academy of Sciences, Grant No. E129952YR0. X. Z. is supported by new faculty startup funding by the Institute of Modern Physics, Chinese Academy of Sciences, by Key Research Program of Frontier Sciences, Chinese Academy of Sciences, Grant No. ZDB-SLY-7020, by the Natural Science Foundation of Gansu Province, China, Grant No. 20JR10RA067 and by the Strategic Priority Research Program of the Chinese Academy of Sciences, Grant No. XDB34000000. J. P. V. is supported by the Department of Energy under Grants No. DE-FG02-87ER40371, and No. DE-SC0018223 (SciDAC4/NUCLEI). A portion of the computational resources were also
provided by Gansu Computing Center.

\end{multicols}
\medline
\begin{multicols}{2}

\end{multicols}
\end{document}